\documentclass[pdflatex,sn-nature]{sn-jnl}

\usepackage{graphicx}%
\usepackage{multirow}%
\usepackage{amsmath,amssymb,amsfonts}%
\usepackage{amsthm}%
\usepackage{mathrsfs}%
\usepackage[title]{appendix}%
\usepackage{xcolor}%
\usepackage{textcomp}%
\usepackage{manyfoot}%
\usepackage{booktabs}%
\usepackage{algorithm}%
\usepackage{algorithmicx}%
\usepackage{algpseudocode}%
\usepackage{listings}%
\usepackage{caption}
\usepackage{newfloat}

\DeclareFloatingEnvironment[
  fileext=loe,        
  placement={!ht},
  name=Extended Data Fig.
]{extfigure}

\DeclareFloatingEnvironment[
  fileext=loe,
  placement={!ht},
  name=Extended Data Table
]{exttable}

\usepackage[nameinlink]{cleveref} 
\crefname{extfigure}{Extended Data Fig.}{Extended Data Figs.}
\Crefname{extfigure}{Extended Data Fig.}{Extended Data Figs.}
\crefname{exttable}{Extended Data Table}{Extended Data Tables}
\Crefname{exttable}{Extended Data Table}{Extended Data Tables}

\newcommand{\aj}{Astron. J.}   
\newcommand{\apj}{Astrophys. J.}   
\newcommand{\apjl}{Astrophys. J. Lett.}   
\newcommand{\apjs}{Astrophys. J. Suppl. Ser.}   
\newcommand{\aap}{Astron. Astrophys.}   
\newcommand{\mnras}{Mon. Not. R. Astron. Soc.}   

\begin{document}
\title[Article Title]{Evidence of triggered star formation in the pillars of creation from JWST observations}


\author[1,2]{\fnm{Jing} \sur{Wen}}
\author*[3]{\fnm{Bingqiu} \sur{Chen}}\email{bchen@ynu.edu.cn}

\author*[1,2]{\fnm{Jian} \sur{Gao}}\email{jiangao@bnu.edu.cn}

\author[4]{\fnm{Jun} \sur{Li}}

\author[5]{\fnm{Ming} \sur{Yang}}

\author[1,2]{\fnm{Biwei} \sur{Jiang}}

\affil[1]{\orgdiv{Institute for Frontiers in Astronomy and Astrophysics}, \orgname{Beijing Normal University}, \orgaddress{\city{Beijing}, \postcode{102206}, \country{People’s Republic of China}}}

\affil[2]{\orgdiv{School of Physics and Astronomy}, \orgname{Beijing Normal University}, \orgaddress{\city{Beijing}, \postcode{100875}, \country{People’s
Republic of China}}}

\affil[3]{\orgdiv{South-Western Institute for Astronomy Research}, \orgname{Yunnan University}, \orgaddress{\city{Kunming}, \postcode{650091}, \state{Yunnan}, \country{People’s Republic of China}}}

\affil[4]{\orgdiv{Center for Astrophysics}, \orgname{Guangzhou University}, \orgaddress{\city{Guangzhou}, \postcode{510006}, \country{People’s Republic of China}}}

\affil[5]{\orgdiv{Key Laboratory of Space Astronomy and Technology}, \orgname{National Astronomical Observatories, Chinese Academy of Sciences}, \orgaddress{\city{Beijing}, \postcode{100101}, \country{People’s Republic of China}}}


\abstract{Stars form in molecular clouds under the influence of their local environments, yet the role of massive stellar feedback in either triggering or suppressing star formation remains a fundamental question in astrophysics. The Pillars of Creation in the Eagle Nebula—sculpted by ionizing radiation and stellar winds from massive stars in the NGC 6611, offer a natural laboratory for investigating this question. Here, we present high-resolution observations of the Pillars of Creation using JWST NIRCam and MIRI, revealing 253 young stellar object (YSO) candidates. These YSO candidates show spatial correlations with the edges of feedback-driven structures, with overdensities along the boundaries. A weak trend of decreasing stellar age with increasing distance from the ionizing source was tentatively observed. There also appears to be an enhancement in the star formation rate  within the past 1\,Myr in this region. Such age and spatial associations suggest that, while the bulk of the YSOs may have formed contemporaneously with the central cluster, a subset could be associated with triggered star formation. The JWST image of intricate structures—including spiral-like disk, bi-reflection nebulae at the tips of Pillar~\uppercase\expandafter{\romannumeral1} and Pillar~\uppercase\expandafter{\romannumeral2} further highlights the complexity of star formation processes.}  





\maketitle


Massive stars play a complex, dual role in the process of star formation. On one hand, their intense radiation ionizes H\,\textsc{ii} regions and disperses molecular clouds through powerful stellar winds, which can inhibit star formation. On the other hand, the same radiation and winds can compress surrounding gas, triggering radiative collapse and initiating the formation of new stars \cite{2012MNRAS.427.2852D,2021SciA....7.9511L}. The balance between these opposing effects—suppression versus triggering—remains an open question in astrophysics. Furthermore, the specific mechanisms of triggered star formation and its overall contribution to the global star formation rate (SFR) remain poorly understood, underscoring the importance of this topic to the broader study of star formation and galaxy evolution.

The Eagle Nebula (M16) is a particularly intriguing star-forming region, providing an ideal environment to investigate these questions. Within the M16, the iconic ``Pillars of Creation" are towering structures of gas and dust sculpted by the feedback from massive stars in the central NGC 6611 cluster \cite{2008hsf2.book..599O}. Such pillar-like formations are characteristic of feedback-driven processes \cite{2012MNRAS.427.2852D}. Early studies using  Hubble Space Telescope (HST) data suggested that dense Evaporating Gaseous Globules (EGGs) within the pillars contain young stellar objects (YSOs), highlighting the pillars as an active star-forming region \cite{1996AJ....111.2349H}. However, follow-up studies using near-infrared (NIR), mid-infrared (MIR), and X-ray observations have raised questions about this interpretation. These studies did not identify a large number of YSO candidates within the Pillars of Creation, for example, \cite{2002A&A...389..513M,2002ApJ...570..749T,2007ApJ...666..321I,2007ApJ...654..347L,2013ApJS..209...32B,2013ApJS..209...31P}. Many EGGs showed little evidence for embedded YSOs or clear signatures of triggered star formation within the Pillars. For such highly obscured star-forming regions, observational limitations have long hindered definitive conclusions.

Infrared observations have proven critical for identifying YSOs through their characteristic infrared excesses, caused by circumstellar dust in their envelopes or disks \cite{2004ApJS..154..363A}. However, interpreting YSOs based solely on infrared photometry can be difficult due to the complex dependence on envelopes or disks geometry and physical properties. A more reliable approach involves combining  NIR and MIR color-magnitude diagrams (CMDs) or color-color diagrams (CCDs) with spectral energy distribution (SED) modeling to identify and classify YSOs \cite{2007ApJ...666..321I}. Nonetheless, given the limited instrumental resolution and depth currently available, identifying YSOs in dense, dust-rich regions remains a significant challenge, as gas and dust obscuration complicates estimates of the true population and spatial distribution of deeply embedded protostars. Moreover, even when YSOs are identified, it is difficult—particularly in small-scale scenarios where star formation is already underway—to distinguish between those formed via feedback-triggered processes and those formed spontaneously \cite{2011EAS....51...45E,2015MNRAS.450.1199D}.Based on the latest observational data, further investigations of these issues may be possible.

The advent of the James Webb Space Telescope (JWST) offers transformative potential for addressing these challenges. With its unprecedented spatial resolution and sensitivity, JWST enables detailed studies of deeply embedded YSOs, revealing their properties and spatial distributions with unparalleled clarity. These capabilities provide a unique opportunity to explore long-standing debates about the role of massive stellar feedback in triggering star formation and to gain new insights into the processes shaping star-forming regions like the Pillars of Creation.

\begin{figure}[htbp]
  \centering
  \includegraphics[width=0.7\textwidth]{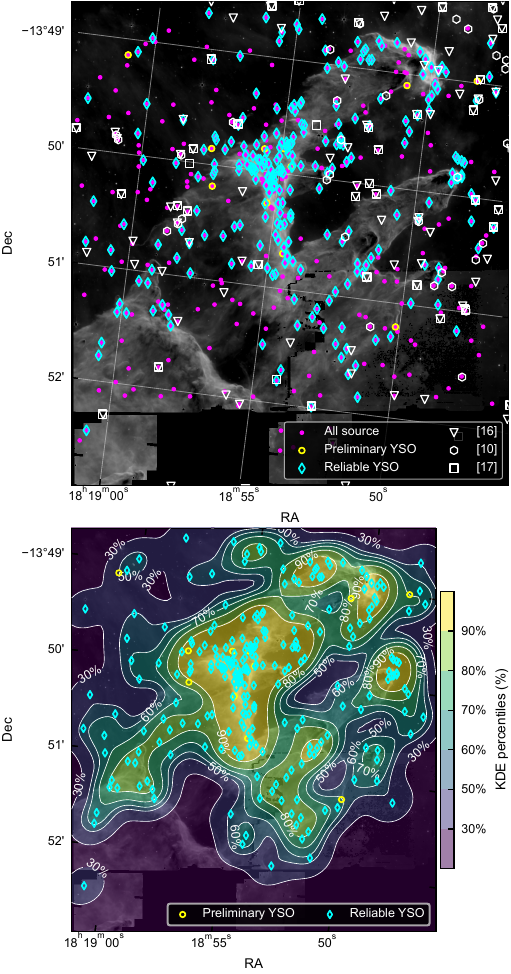}
  \caption{Spatial distribution of the identified YSO candidates. The background image in both panels is the JWST F770W image. Upper panel: Pink points represent initial sources that satisfy at least one infrared color-selection criterion (See Methods, Extended Data Table~1). Hollow scatter points correspond to different Fflag values: yellow circles indicate preliminary candidates ($\mathrm{Fflag} = 1$; i.e., sources meeting the quality criteria of only one fitting method), while cyan diamonds represent more reliable YSO candidates ($\mathrm{Fflag} = 2$; i.e., sources meeting the criteria of both fitting methods). White markers denote YSOs and X-ray-selected pre-main sequence (PMS) stars reported in previous studies, with different symbols indicating different catalogs, inverted triangles: \cite{2012ApJ...753..117G}, hexagons: \cite{2013ApJS..209...31P}, and squares: \cite{2014ApJ...787..108G}. Lower panel: Kernel Density Estimate (KDE) of the reliable YSO candidates ($\mathrm{Fflag} = 2$). White contours mark quantiles of the KDE values, and colors fill the intervals between adjacent quantile levels.}
  \label{fig:1}
\end{figure}

Using multi-band photometry from  JWST, combined with SED fitting and CMDs (See Methods, Extended Data Table~1, Extended Data Figure~1, Extended Data Figure~2), we identified a total of 264  YSO candidates, among which 253 are classified as reliable YSO candidates ($\mathrm{Fflag} = 2$), and 11 are considered preliminary candidates ($\mathrm{Fflag} = 1$). Due to the stringent selection criteria applied in this study, our candidate list prioritizes purity over completeness. Figure~\ref{fig:1} illustrates the spatial distribution and Kernel Density Estimation (KDE) contours of the YSO candidates alongside the positions of previously identified YSOs and X-ray sources in the M16. The number of YSO candidates identified in this study significantly exceeds that reported in previous investigations of this region. \cite{1982MNRAS.199P...9W} identified only eight NIR sources near the pillars. Using MIR observations from Space Observatory camera (ISOCAM), \cite{1998A&A...333L...9P} found only one embedded source associated with an EGG. \cite{2002A&A...389..513M} analyzed 73 EGGs using Infrared Spectrometer And Array Camera (ISAAC) JHK imaging with a resolution of $\sim 0.35^{\prime \prime}$ and suggested that only about 15\% of EGGs might contain low-mass YSOs. Subsequent studies by \cite{2012ApJ...753..117G,2013ApJS..209...31P, 2014ApJ...787..108G} expanded and studied the sample of young stars detected via X-ray emission and infrared data (shown as white markers in the top panel of Figure~\ref{fig:1}). However, none of these datasets revealed a significant population of YSOs within the pillars.

These studies likely represent a lower limit, constrained by the limited sensitivity and resolution of earlier observations, which were insufficient to detect and resolve deeply embedded or low-mass stars—particularly in the very bright rim along the base of Pillar I. These limitations likely led prior studies to underestimate the star formation activity within the pillars, resulting in the conclusion that star formation in this region was relatively low \cite{2007ApJ...666..321I}. In contrast, the significantly larger number of YSO candidates identified in JWST observations provides evidence for active star formation occurring within the pillars.

The spatial distribution of YSO candidates shows a clear clustering pattern along the edges of the dust and gas pillars, with the highest concentration near the bright-rimmed cloud of Pillar~\uppercase\expandafter{\romannumeral1} base. Such spatial configurations are a hallmark of triggered star formation \cite{2012MNRAS.421..408T,2012ApJ...755...71K,2016MNRAS.461.2502Y}. This observation also aligns with numerical simulations \cite{2012MNRAS.427.2852D,2013MNRAS.431.1062D}, which show that triggered star formation predominantly occurs near pillar structures and the boundaries of expanding bubbles. 

However, as \cite{2015MNRAS.450.1199D} point out after surveying 67 observational studies on triggered star formation, the effectiveness of observational criteria is limited. Relying on a single indication of triggering does not suffice to prove that star formation was triggered. This is because stellar feedback not only triggers new star formation but also redistributes stars that would have formed regardless. Consequently, star clusters observed near feedback-driven shells or pillar-like structures may consist of a mixture of triggered stars and spontaneously formed stars. If a given region exhibits additional evidence beyond mere geometric association, then the likelihood that its star formation was genuinely triggered increases. We proceed with the following discussion under this premise.

\begin{figure}
\centering
\includegraphics[width=0.6\textwidth]{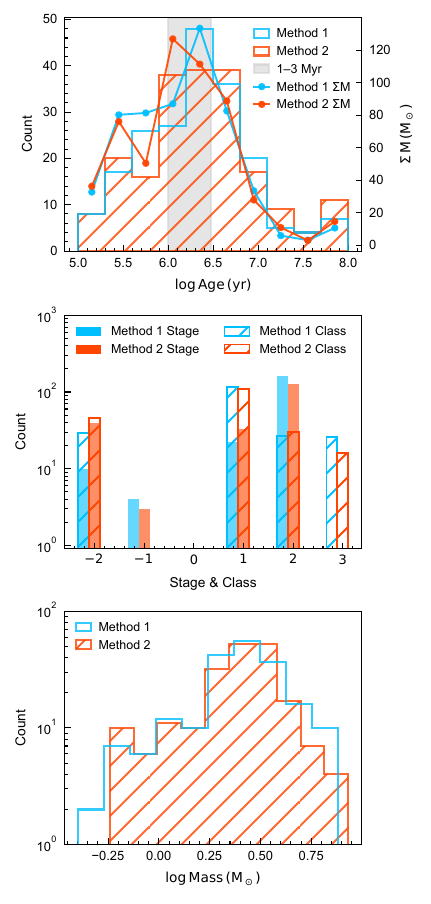}
\caption{Parameter distributions of the YSO candidates. Orange and cyan colors represent results obtained from two different SED fitting methods: Bayesian Model Comparison (Method1, BMC, see Methods) and Comprehensive $\chi^{2}$ Fitting (Method2, CCF). All three panels include only those YSOs with $\mathrm{Fflag} = 2$ (i.e., sources that meet the criteria of both fitting methods) and $\mathrm{Aflag} = 0$ (i.e., sources that also satisfy the isochrone selection requirement, lying between the zero-age main sequence (ZAMS) and the $10^{5}$\,yr isochrone). The top panel shows the age histogram, with the overlaid line indicating the total stellar mass within each age bin. The gray shaded region marks previous estimates for the age of NGC6611 ($\sim$1–3\,Myr; \cite{1993AJ....106.1906H,2006A&A...457..265D}). Middle panel: Classification results of Stage and Class. Sources with $\mathrm{Stage\ (Class)} = -2$ are defined as ambiguous in our scheme. Other values follow the model parameters defined in \cite{2024ApJ...961..188R}: $\mathrm{Stage={0,1,2,3,-1}}$, with Stage 0, I, II, and III representing successive evolutionary phases approaching the main sequence, and $\mathrm{Stage}=-1$ denoting ``does not comport / incomplete." Similarly, $\mathrm{Class={0,1,2,3,4,-1}}$, with Class 0, I, Flat, II, and III corresponding to observational classes of the models based on the infrared spectral index, and $\mathrm{Class}=-1$ denoting ``no spectral index / unclassified / incomplete." The bottom panel shows the mass histogram. Note that the Y-axis scale differs among the three panels.}

\label{fig:4}
\end{figure}

\begin{figure}
\centering
\includegraphics[width=0.9\textwidth]{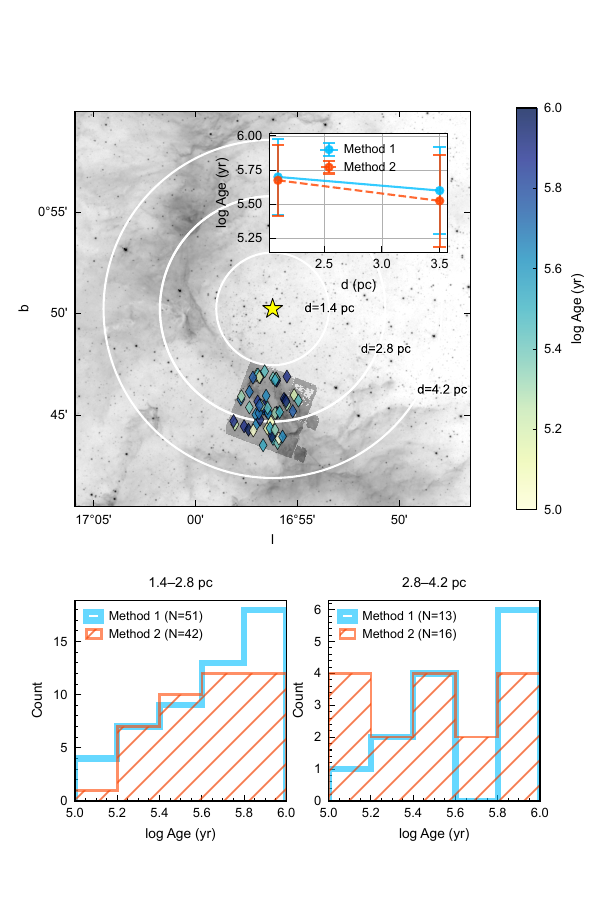}
\caption{The spatial distribution of stellar ages. In the top panel, the JWST/F770W image (shown in a darker shade) is overlaid on the Spitzer/IRAC3 background, covering a smaller region. The central yellow star symbol marks the position of the primary ionizing source, W205. A distance of 1.74\,kpc \cite{2019ApJ...870...32K} is adopted in this analysis. White concentric circles, drawn at intervals of 1.4\,pc, mark the distance $d$ from the ionizing source. The color of the scatter points in the main panel represents stellar ages derived using Method2, but only sources with $\mathrm{Fflag} = 2$, $\mathrm{Aflag} = 0$, and ages $\leq 1$\,Myr are included (i.e., sources that meet the criteria of both SED fitting methods and also satisfy the isochrone selection requirement). The inset in the upper-right corner shows the variation of the median YSO age across radial bins with distance from the ionizing source. The error bars denote the standard deviation within each distance bin, $s=\sqrt{\frac{1}{N-1}\sum_{i=1}^{N}(x_i-\bar{x})^2}$. The the bottom panels show the age distributions within the radial bins.}

\label{fig:3}
\end{figure}

\begin{figure}
\centering
\includegraphics[width=1\textwidth]{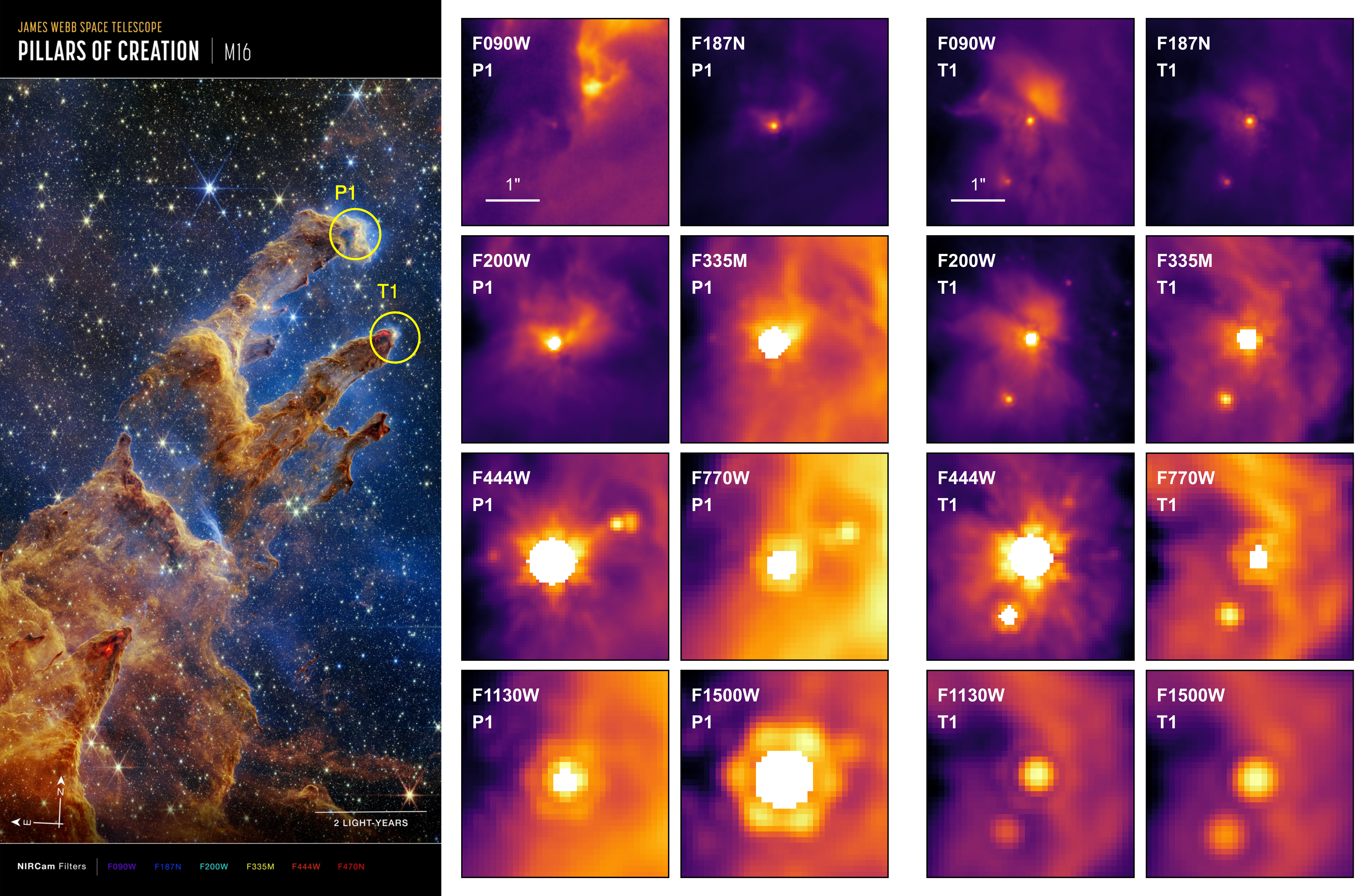}
\caption{The Pillars of Creation observed by JWST. Left panel shows the JWST/NIRCam infrared press image of the Pillars of Creation (M16; observed on 14 August 2022; Program 2739, PI: K. Pontoppidan; credit: NASA, ESA, CSA, and STScI; image processing: J. DePasquale (STScI), A. M. Koekemoer (STScI), and A. Pagan (STScI) (\href{https://www.nasa.gov/universe/nasas-webb-takes-star-filled-portrait-of-pillars-of-creation/}{NASA Press Release: https://www.nasa.gov/universe/nasas-webb-takes-star-filled-portrait-of-pillars-of-creation/}). The positions of T1 and P1 are marked on the image. The middle and right panels show JWST images of two well-known and extensively studied YSOs in the Pillars of Creation, P1 and T1, across eight bands: from top-left to bottom-right, F090W, F187N, F200W, F335M, F444W, F770W, F1130W, and F1500W. The JWST images reveal their complex structures and nearby companion sources. Each subimage is a square of 4 $\times$ 4 arcsec. In the F090W image of each source, a scale bar indicates the length corresponding to 1 arcsec.}
\label{fig:RGB}
\end{figure}

We derived key physical parameters for the YSO candidates from SED fitting, including luminosity, effective temperature, evolutionary stage, and classification. Two different fitting criteria were adopted as independent methods, and we calculated weighted average values of the parameters rather than directly using those from the best-fit models. Masses and ages were subsequently determined by fitting the derived effective temperatures and luminosities to the MIST isochrones \cite{2016ApJS..222....8D,2016ApJ...823..102C} (See Methods, Extended Data Figure~3). The YSOs located between the zero-age main sequence (ZAMS) and the $10^{5}$\,yr isochrone are flagged with $\mathrm{Aflag} = 0$. Among the candidates with $\mathrm{Fflag} = 2$, a total of 198 and 201 sources satisfy $\mathrm{Aflag} = 0$ under Method 1 and Method 2, respectively. The statistical distributions of  parameters are shown in the Figure~\ref{fig:4}.

Previous studies have estimated the ages of the main members of NGC 6611 to be around 1 to 3\,Myr \cite{1993AJ....106.1906H,2006A&A...457..265D}. \cite{2023A&A...670A.108S}  suggested the existence of two distinct stellar populations in M16: a younger population with an age of $1.3 \pm 0.2$\,Myr and an older population with an age of $7.5 \pm 0.4$\,Myr. Most stars in the cluster center appear to have formed within the past $\sim$3\,Myr, indicating that a major star formation event shaped the cluster. However, the presence of older members suggests that the star formation history (SFH) of M16 is not characterized by a single episode, but rather involves a complex, multi-stage process. The OB stars near the Pillars of Creation are generally associated with the younger population. As shown in the age histogram in Figure~\ref{fig:4}, if the peak is interpreted as representing the period of most active star formation, it roughly coincides with previous estimates for the age of NGC 6611 (the shaded region).

If the YSOs within the pillars were triggered, then under the conservative constraints, they would need to be younger than $1.3$\,Myr and the timescale required for the feedback to reach their current positions would also need to be taken into account. Although  many YSO candidates in our sample have ages older than 1\,Myr, suggesting that their formation likely predated the arrival of feedback from massive star, a substantial population of very young candidates (ages $\leq 1$\,Myr) are also present. Among the YSOs flagged with $\mathrm{Fflag} = 2$ and $\mathrm{Aflag} = 0$, we identify 64 and 58 sources with ages $\leq 1$\,Myr based on Methods 1 and 2, respectively. As shown in the middle panel of Figure~\ref{fig:4}, SED-based classifications indicate that many of these sources are in early evolutionary stages.

To make a rough estimate of the feedback timescale, according to the Spitzer solution \cite{1978ppim.book.....S} (see Method) , the ionization front would reach the tips of the pillars ($\sim$1.5\,pc) in about 0.1\,Myr. The time required to reach the bottom of the pillar structures ($\sim$3.5\,pc) would be approximately 0.6\,Myr, or about 0.5\,Myr using the  Hosokawa $\&$ Inutsuka analytical solution \cite{2006ApJ...646..240H}. If the YSOs in the pillars were triggered, we would expect to observe a sufficiently young population may exhibiting a trend of decreasing age with increasing distance from the ionizing source. Figure~\ref{fig:3} shows the spatial distribution of sufficiently young YSOs, selected with ages $\leq 1$\,Myr. In the point–line diagram, we tentatively observe a weak trend. The bottom panel displays the age distributions within each radial bin, which also appear to show a weak trend toward younger ages with increasing distance.

However, it is important to note that the uncertainties in the age estimates are large. The error bars in the point–line diagram of Figure~\ref{fig:3} likewise show large standard deviations, reminding us that this result should be interpreted with caution. Observations and simulations of other star-forming regions have reported difficulties in detecting clear age gradients \cite[e.g.,][]{2013MNRAS.431.1062D,2020ApJ...897...74Z}. Traditionally, age gradient analysis has been employed to investigate the formation history of star-forming regions and to distinguish between triggered and spontaneous star formation. Recent studies have shown that the presence of an age gradient alone is insufficient to confirm triggering, as similar gradients can arise in both triggered and spontaneous star formation scenarios \cite{2013MNRAS.431.1062D}. In addition, the projected distances do not necessarily reflect the true three-dimensional separations between the YSOs and the ionizing source.  Within this relatively small region—where even the determination of stellar ages is inherently uncertain—it is likely to be even more difficult to robustly confirm the presence of an age gradient.

To explore whether stellar feedback influences the local SFR, we performed a simple SFR estimate using a subset of 64 YSO candidates from our sample, each with an age $\leq$ 1\,Myr as determined by Method 1. The estimated SFR is approximately $2.4 \times 10^{-4}\,M_{\odot}\,\mathrm{yr^{-1}}$. Assuming that these YSOs are distributed over an area of about 3.5\,pc$^{2}$, the corresponding  $\Sigma_{\mathrm{SFR}}$ is $\sim 6.9 \times 10^{-5}\,M_{\odot}\,\mathrm{yr^{-1}\,pc^{-2}}$. \cite{2010ApJ...724..687L} proposed a  relation between the SFR and the mass of dense gas in molecular clouds. According to this relation, the expected SFR $= 4.6 \times 10^{-8} M_{0.8}$ $M_{\odot}\,\mathrm{yr^{-1}}$, where $M_{0.8}$ is the cloud mass above an extinction threshold of $A_{\rm{K}} \sim 0.8$\,mag. \cite{1999A&A...342..233W} estimated the total mass of the three main pillars to be approximately 200\,$M_{\odot}$, while \cite{2023AJ....166..240K} reported the total mass of pillars to be around 300\,$M_{\odot}$. Applying these values to the \cite{2010ApJ...724..687L} relation yields an expected SFR in the range of $0.9 \sim 1.4 \times 10^{-5}\,M_{\odot}\,\mathrm{yr^{-1}}$. Our estimated SFR is  higher than previous values. Given that our YSO sample is incomplete, this estimate likely represents a lower limit, suggesting that star formation in this region seems have been more active over the past $\sim$1\,Myr than previously inferred. It is important to note, however, that although the \cite{2010ApJ...724..687L} relation adopts a threshold of $A_{\rm{K}} \sim 0.8$\,mag to connect the SFR with the dense regions rather than the entire cloud, our data still cover only a limited portion of M16. This area corresponds to a potentially more active subregion of star formation, and therefore the derived SFR may appear higher when compared with global SFR–gas relations that represent averages across the entire cloud. Observational data with a broader spatial coverage are still required to determine whether the elevated SFR is indeed caused by feedback effects.

Within this limited region, another perspective we can analyze is the variation of the star formation over time. As shown in the age histogram in Figure~\ref{fig:4}, the number of YSOs on the left side of the peak exceeds those on the right side across both fitting methods. The overlaid line, representing the total mass within each bin, shows a similar trend. This could indicate a slight increase in the SFR following the formation of the massive O- and B-type stars, potentially reflecting the role of stellar feedback in triggering star formation within this region.

These findings may imply that we are witnessing triggered star formation. Nevertheless, we remain cautious in this interpretation, as the derived ages may carry substantial uncertainties arising from factors such as YSO variability, extinction, model assumptions, and the SED-fitting strategy—especially when only a limited number of photometric bands are available. For example, as shown in the middle panel of Figure~\ref{fig:4}, many of  YSOs fall into the ambiguous classification category of $-2$. The available photometric data alone are insufficient to constrain YSO ages with high precision. While it is possible that some of these sources are indeed triggered stars, a considerable degree of uncertainty remains.

Figure~\ref{fig:4} also presents the mass histogram of the YSO candidates. At the high-mass end, the mass distributions obtained from different methods are highly consistent. For the more reliable candidates ($\mathrm{Fflag} = 2$), the lower mass limit is approximately 0.3\,$M_{\odot}$, with an average YSO mass of about 2.7\,$M_{\odot}$. This result differs from earlier studies. Previous work based on ISOCAM suggested a relatively low level of ongoing low-mass star formation in this region \cite{1998A&A...333L...9P}. Using the improved resolution and sensitivity of Spitzer, \cite{2007ApJ...666..321I} confirmed this inference by detecting no additional mid-infrared sources in the Pillars of Creation, down to flux limits of $\sim$1 mJy at [3.6] and 5 mJy at [8.0], corresponding to $\sim$0.4\,$M_{\odot}$. This may be because, in the Pillars of Creation, the high concentration of YSO candidates appears along a bright rim at the base of Pillar I. Due to the limited resolution of previous telescopes, some YSOs may have been blended with the nebular background. Although we did identify some previously undetected low-mass YSOs, the mass distribution shown in the bottom  panel of Figure~\ref{fig:4} indicates that our YSO sample remains incomplete. Identifying low-mass, low-luminosity YSOs remains a challenge.

At the tips of Pillar~\uppercase\expandafter{\romannumeral1} and Pillar~\uppercase\expandafter{\romannumeral2}, high-density molecular cores have been identified, consistent with the expected compression induced by the surrounding H\,\textsc{ii} region, bipolar outflow associated with embedded protostars have also been detected. Two sources—P1 and T1—have been extensively studied as representative YSO candidates at the pillar tips \cite[e.g.,][]{2002ApJ...565L..25S, 2002ApJ...568L.127F, 2002ApJ...570..749T,2007ApJ...666..321I,2015MNRAS.450.1057M}. However, as illustrated in Figure~\ref{fig:RGB}, both P1 and T1 are saturated in multiple photometric bands, which limits our ability to perform detailed SED fitting and parameter estimation. Despite this limitation, inspection of the high-resolution images still provides valuable insights into their structures, as discussed below.

JWST observations reveal that P1 is likely a composite of multiple sources rather than a single YSO. As shown in the middle panel of Figure~\ref{fig:RGB}, two distinct sources are resolved within approximately $1^{\prime \prime}$ in the F444W band.  The two nearby sources to the right are clearly resolved in the F444W band and are also visible in the F770W and F1500W bands, though they cannot be fully separated. However, the source to the left is only detected in the F444W band. The circumstellar  disk surrounding P1, visible in the F187N and F200W images, displays spiral-like structures. The structure on the right extends over $1^{\prime \prime}$. Considering the distance to M16 is approximately 1.6–1.8\,kpc \cite{2019ApJ...870...32K,2020A&A...633A..51Z,2021MNRAS.504..356D}, this corresponds to a physical scale of 1600–1800\,AU. Meanwhile, the left side appears slightly shorter and fainter across all bands. However, the image could also represent a large-scale accretion stream or an outflow.  Additionally, a dark lane appears at the center of the F187N and F200W images, which may correspond to an inclined disk, this feature could potentially be interpreted as a highly inclined YSO, similar to T1, which we will discuss next. Overall, it is difficult to determine the structure based solely on the current imaging data.

Another well-studied YSO, T1, is located at the tip of Pillar~\uppercase\expandafter{\romannumeral2}. JWST observations show that T1 is a high-inclination YSO. In the F090W, F187N, and F200W images, two asymmetric fan-shaped regions are visible around T1. These bi-reflection nebulae, separated by a dark lane, are characteristic of high-inclination protoplanetary disks. In the F090W image, the northern region appears brighter, while the southern region is fainter. However, this pattern is nearly reversed in the F200W band. This phenomenon may be explained by a misaligned inner disk or a warped disk, although further investigation is required. Recent studies suggest that infrared flux reversals are a potential signature of disk warping \cite{2025A&A...698A.146K}.  The southern fan-shaped region extends up to $3^{\prime \prime}$ in the F200W image, corresponding to a physical distance of approximately 4000 to 5000 AU—about twice the extent of the northern side. In the F187N image, the nebulae appear smaller. In the F090W band, the width of the central dark lane is around $0.2^{\prime \prime}$ (approximately 300 to 400 AU). In all bands, a nearby star in the southern region may influence the observations of T1.

In summary, our findings suggest that some of the YSOs we identified exhibit several characteristics commonly associated with triggered star formation, including their spatial distribution and age properties. Star formation activity in this region over the past 1\,Myr also appears to have been more vigorous than previously estimated. However, confirming whether these are truly triggered stars based solely on photometric data remains challenging, and will require further observational and theoretical investigation. It is clear that additional data and supporting evidence are needed.  Notably, the observations of the spiral-disk structure of P1, the bi-reflection nebula structure of T1, and their associated companions underscores the extraordinary potential of JWST for advancing our understanding of YSO structure and evolution. 

\clearpage
\section*{Methods}\label{sec:method}

\subsection*{Data} 
For this study, we utilized publicly available JWST observations of the Eagle Nebula (M16) from the Director’s Discretionary (DD) program (ID: 2739; PI: Klaus Pontoppidan). These observations included NIRCam filters (F090W, F187N, F200W, F335M, and F444W) and MIRI filters (F770W, F1130W, and F1500W), providing comprehensive coverage of the main structures of the Pillars of Creation. The fields of view for the Pillars of Creation are 7.4 $\times$ 4.4 arcmin$^{2}$ (for NIRcam) and 4.3 $\times$ 3.8 arcmin$^{2}$ (for MIRI), respectively. 

Photometric data for this study were sourced from \cite{2024ApJ...968L..26L}, who utilized point spread function (PSF) photometry performed with the STARBUG~\textsc{ii}  \cite{2023ascl.soft09012N}. This tool is specifically optimized for analyzing JWST observations in dense and complex stellar fields. Approximately 400,000 sources were extracted from this region by \cite{2024ApJ...968L..26L}. For detailed descriptions of the photometric processing, we refer readers to their work.

\subsection*{Identification of YSO Candidates}
We identified YSO candidates using a combination of CMDs, SED fitting, and visual inspection. Multiple CMDs were constructed for this analysis, as shown in the Extended data Figure~1. The M16 suffers from severe extinction, which reddens all sources. Despite this, the CMDs clearly show two distinct groups of sources. In the right region of the diagrams, sources appear significantly redder, likely due to circumstellar material surrounding young stars, means that these CMDs can be readily used as tools to find and classify YSOs. We defined four color criteria to isolate YSO candidates. Extended Data Table~1 lists the number of sources meeting each criterion (Cflag). A total of 485 sources were selected for further analysis. These sources met at least one color criterion and had photometric data in at least four bands. Among the 485 sources with $\mathrm{Cflag} \geq 1$, 215 have photometry in 4 bands, 148 in 5 bands, 66 in 6 bands, 37 in 7 bands, and 19 in all 8 bands. This multi-band data provided the foundation for subsequent classification and analysis of YSO candidates.

Next, we utilized the YSO SED models developed by \cite{2017A&A...600A..11R}, with updates from \cite{2024ApJ...961..188R} that include model-convolved JWST photometry and other parameters. These models cover a broad range of YSO evolutionary phases, from deeply embedded protostars surrounded by envelopes and accretion disks to more evolved PMS stars. For our analysis, we constrained the distance of all sources to 1.6–1.8\,kpc \cite{2019ApJ...870...32K,2020A&A...633A..51Z,2021MNRAS.504..356D} and adopted an extinction range of $A_{V}=0.001$ to $100$\,mag to account for the high dust obscuration in the M16.

Previous studies often used the “spubhmi” model group for SED fitting. This model group, as defined by \cite{2017A&A...600A..11R}, is comprehensive, incorporating components such as a central star, circumstellar envelope, disk, bipolar cavity with up to 12 parameters. While this complexity enables detailed fitting, it can also reduce the reliability of derived parameters when observational data are limited to 4–8 photometric bands. To mitigate this issue, we implemented two distinct methods to determine the best-fitting model set and the obtain parameters:

1. Bayesian Model Comparison (BMC): Following the methodology of \cite{2017A&A...600A..11R}, we calculated the likelihood $P(D \mid M)$ for each model set and normalized these values to derive a score. The score was defined as the $P(D \mid M)$ of the model set normalized by the average $P(D \mid M)$ across all sets. The model set with the highest score was selected, and the model with the minimum $\chi^2$ value within that set was chosen as the best-fitting model.

2. Comprehensive $\chi^2$ Fitting (CCF): We fit the observational data across all 18 model sets from \cite{2017A&A...600A..11R}, spanning the star-only “s-s-i” group to complex “spubhmi” group. The model with the lowest $\chi^2$ value across all sets was selected as the best-fitting model.

For $\chi^2$ calculations, we utilized the Python package \texttt{SEDfitter} \cite{2007ApJS..169..328R}. The $\chi^2$ formula used is as follows:
\begin{equation}
\chi^2 = \sum_{i=1}^{N} \left( \frac{\langle \log_{10}[F_\nu(\lambda_i)] \rangle - \log_{10}[M_\nu(\lambda_i)]}{\sigma(\langle \log_{10}[F_\nu](\lambda_i) \rangle)} \right)^2,
\label{eq1}
\end{equation}
where $\langle \log_{10}[F_\nu(\lambda_i)] \rangle$ represents the observed  flux at a given wavelength $\lambda_i$, $\sigma(\langle \log_{10}[F_\nu](\lambda_i) \rangle)$ is the flux uncertainty, and $\log_{10}[M_\nu(\lambda_i)]$ is the extincted and scaled model flux. 
The $P(D \mid M)$, and the score used in the BMC method follow the formalism outlined in \cite{2017A&A...600A..11R} and are briefly expressed as:
\begin{equation}
P(D \mid M) \propto \frac{N_{\text{good}}}{N},
\end{equation}
where $N_{\text{good}}$ is the number of models within the model set $M$ that meet the goodness-of-fit criterion, and $N$ is the total number of models in the set. The goodness-of-fit criterion is defined as:

\begin{equation}
\chi^2 - \chi^2_{\mathrm{best}} \leq 9\,n_{\mathrm{data}}
\label{eq:chi}
\end{equation}

where $\chi^2_{\text{best}}$ is the lowest $\chi^2$ value among all model sets, and $n_{\text{data}}$ is the number of photometric data points used in the fitting.

Using the results from the BMC and CCF methods, many of the best-fitting models were simpler, containing fewer components and consequently reducing the number of free parameters. When the observed SED can be adequately reproduced by multiple models, those with fewer free parameters are preferred due to their simplicity and reliability. Extended Data Figure~2 illustrates an example of an SED fit for a YSO candidate using all two fitting methods. The title of each panel displays the best-fitting model set name, fitting scores, and associated variables, while different colors (red, blue) represent the best-fitting models obtained through the BMC, CCF methods, respectively.

After obtaining fitting results for all 485 sources, we visually inspected the results and excluded 37 sources that the observed SEDs could be well-reproduced by simple models containing only a central star. These 37 sources were flagged as Sflag=1. Next, we classified reliable YSO candidates by selecting sources for which the best-fitting models from both two methods satisfied $\chi^{2}/n_{\text{data}} \leq 10$.  This criterion ensures that only candidates with adequately low $\chi^2$ values are considered ``good" fits. This subset, consisting of 253 sources, was assigned a reliability flag of $\mathrm{Fflag} = 2$. For sources where only one of the two methods satisfied $\chi^{2}/n_{\text{data}} \leq 10$, we assigned a flag of $\mathrm{Fflag} = 1$, resulting in an additional subset of 11 sources.

\subsection*{Physical properties of YSO Candidates} 

 Based on the aforementioned two methods, we may obtain two different best-fit models derived from two distinct model sets. Rather than directly adopting the parameters of these models, we employ the $\chi^2$-weighted parameter averaging approach, as described in \cite{2013ApJS..209...31P,2016ApJ...825..125P}, to determine the final parameters. When a model set is identified as the best-fit modelset  by any one of the two methods, all parameters used in the final $\chi^2$-weighted average are exclusively drawn from that specific set. Specifically, for the $k$ models that satisfy the goodness-of-fit criterion given by Equation~\ref{eq:chi} and are classified as being above the main sequence (using the updated parameter from \cite{2024ApJ...961..188R}, where ``$Above MS = 1$" serves as a filtering criterion), we assign a $\chi^2$-weighted probability to each model using  

\begin{equation}
    P_k = P_n e^{-\chi_k^2 / 2}
\label{eq:p}
\end{equation}
where $P_n$ is chosen such that $\sum_k P_k = 1$. We adopt the same approach as \cite{2013ApJS..209...31P}, using Equation~\ref{eq:p} to classify each YSO candidate according to the probability distribution of its evolutionary stage. The Stage parameter in \cite{2024ApJ...961..188R} is defined as $\mathrm{Stage={0,1,2,3,-1}}$: Stage 0, Stage I, Stage II, and Stage III correspond respectively to evolutionary phases approaching the main sequence, while $\mathrm{Stage}=-1$ denotes “does not comport with the definition / incomplete." Similarly, the observational class (``Class") is defined as $\mathrm{Class={0,1,2,3,4,-1}}$: Class 0, Class I, Flat spectrum, Class II, and Class III correspond respectively to observational classes of the models based on the infrared spectral index, while $\mathrm{Class}=-1$ denotes ``no spectral index / unclassified / incomplete". We apply the following criterion: if the probability $\geq 0.67$, that a source belongs to Stage (Class) it is assigned to that Stage (Class). Otherwise, its evolutionary Stage is considered “ambiguous” and labeled as ``$\mathrm{Class (Stage)} = -2$". For the other parameters, we perform calculations,

\begin{equation}
    \text{Parameter}_{\text{SED}} = \sum_k P_k \text{Parameter}_k
\end{equation}

This approach is applied to two methods to obtain the final two sets of parameters. Based on the derived temperatures and luminosities, we perform isochrone fitting using the MIST \cite{2016ApJS..222....8D,2016ApJ...823..102C}, adopting the KDTree method to determine two sets of masses and ages. The adopted isochrones span an age range from $10^{5}$\,yr (the lower limit of the MIST isochrones) to $10^{8}$\,yr. Extended Data Figure~3 presents the isochrone-fitting results. We label sources in the H–R diagram as $\mathrm{Aflag} = 0$ if they lie between the ZAMS and the $ 10^{5}\,\mathrm{yr}$ isochrone, as $\mathrm{Aflag} = -1$ if they fall below the ZAMS, and as $\mathrm{Aflag} = 1$ if they are above the $10^{5}\,\mathrm{yr}$ isochrone. In the analysis and discussion of the main text, we use only those sources with $\mathrm{Aflag} = 0$.

\subsection*{Feedback Timescale}
Numerous studies have explored the evolution of H\,\textsc{ii} regions, and many analytical solutions are now widely used (e.g., \cite{1977ApJ...218..377W,2009ApJ...703.1352K,2016ApJ...819..137K,1978ppim.book.....S,2006ApJ...646..240H}). To estimate the feedback timescale and thereby identify which sources may have been triggered, we performed a simple calculation, adopting the analytical solutions of \cite{1978ppim.book.....S} and \cite{2006ApJ...646..240H}, as detailed below. The initial Strömgren radius \cite{1939ApJ....89..526S} is given by,
\begin{equation}
R_{\mathrm{S}} = \left( \frac{3 S_{*}}{4 \pi n_{0}^{2} \alpha} \right)^{1/3},
\end{equation}
where $S_{*}$ is the ionizing photon production rate of the central star, $n_{0}$ is the number density of the surrounding medium,  and $\alpha \sim 2.7 \times 10^{-13}\,\mathrm{cm^{3}\,s^{-1}}$ is the case B recombination coefficient. The classical Spitzer solution \cite{1978ppim.book.....S} is given by:
\begin{equation}
R_{\mathrm{Sp}}(t) = R_{\mathrm{S}}
\left[1 + \frac{7}{4} \frac{c_{\mathrm{i}} t}{R_{\mathrm{S}}} \right]^{\tfrac{4}{7}},
\label{eq:sp}
\end{equation}
Here, $R_{\mathrm{Sp}}(t)$ is the radius of the ionization front as a function of time,  and $c_{\mathrm{i}}$ is the sound speed in the ionized gas. The analytical solution of Hosokawa \& Inutsuka \cite{2006ApJ...646..240H} is given by:
\begin{equation}
R_{\mathrm{H\,I}}(t) = R_{\mathrm{S}}
\left[1 + \frac{7}{4} \sqrt{\frac{4}{3}} \frac{c_{\mathrm{i}} t}{R_{\mathrm{S}}} \right]^{\tfrac{4}{7}}.
\label{eq:HI}
\end{equation}
The primary ionizing source in M16 is generally identified as W205 \cite{1996AJ....111.2349H}, which is estimated to have a mass of approximately $75$--$80\,M_\odot$ and a spectral type of O3--O5\,V \cite{1993AJ....106.1906H,2005A&A...437..467E,1995ApJ...454..151M,2000A&A...358..886B}.
 We adopt  $ S_{*} = 1.0 \times 10^{50}$\,s$^{-1}$ and a hydrogen density of $n_0 = 3 \times 10^3$\,cm$^{-3}$\cite{1996AJ....111.2349H,2015MNRAS.453.1324B}. These parameters yield an initial Strömgren radius of approximately 0.7\,pc. Assuming $c_{\mathrm{i}} = 10\,\mathrm{km\,s^{-1}}$, the time required for the ionization front to reach the tip of the pillar (approximately 1.5\,pc) is estimated to be $\sim 0.11$\,Myr and $\sim 0.09$\,Myr based on Equation~\ref{eq:sp} and Equation~\ref{eq:HI}, respectively. The time to reach 3.5\,pc is approximately 0.6\,Myr and 0.5\,Myr from the two equations, respectively. If a higher speed of $c_{\mathrm{i}} = 12.85\,\mathrm{km\,s^{-1}}$ is adopted, the corresponding times to reach 1.5\,pc decrease to $\sim 0.09$\,Myr and $\sim 0.07$\,Myr, while the times to reach 3.5\,pc become $\sim 0.5$\,Myr and $\sim 0.4$\,Myr. Overall, this represents a simplified estimate.
 
\section*{Extended Data}\label{ed}

\begin{extfigure}[t]
\centering
\includegraphics[width=0.85\textwidth]{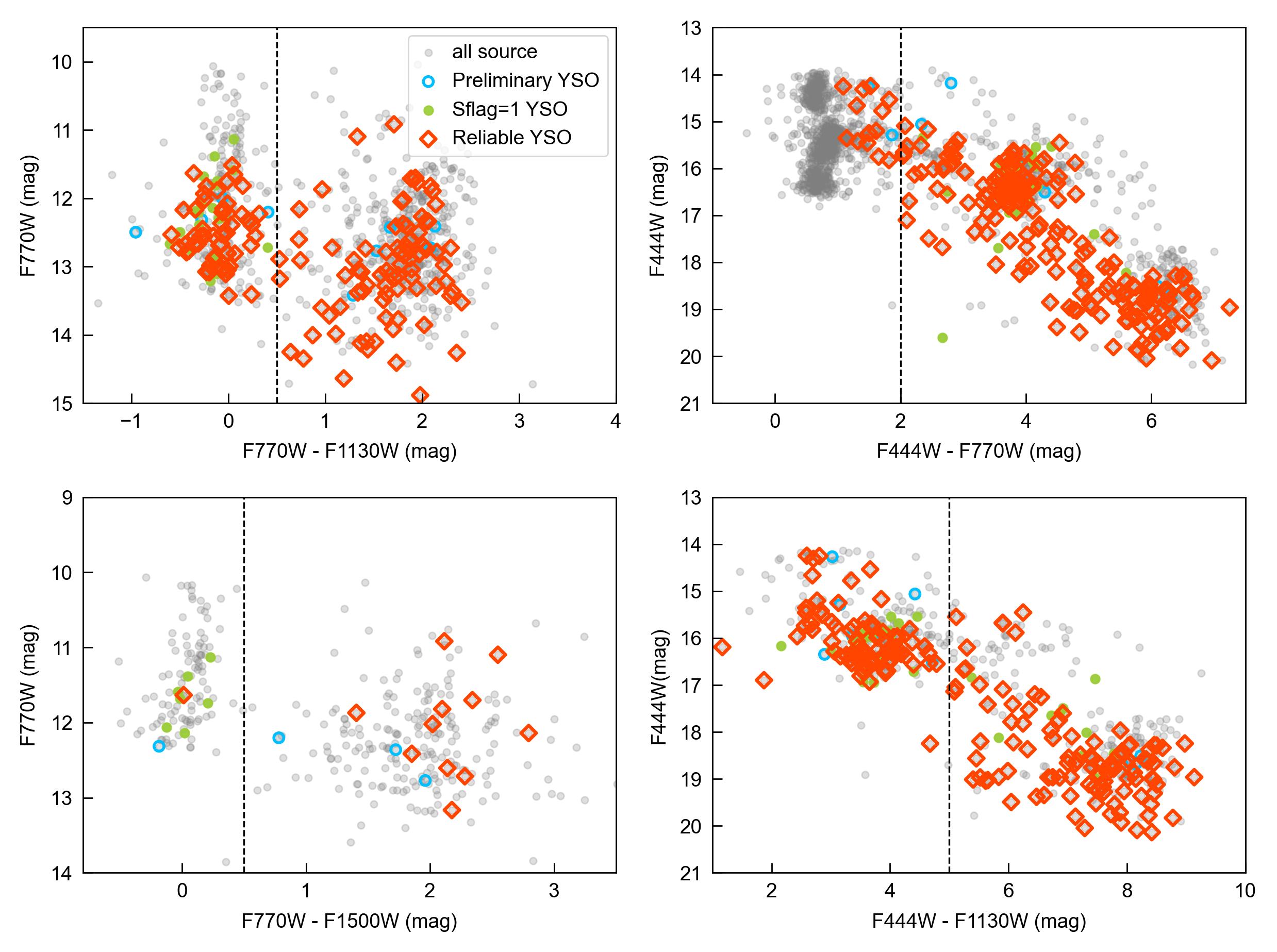}
\caption{Colour-magnitude diagrams (CMDs) of YSO candidates. Each panel shows a different combination of JWST filters. The vertical black lines in the four panels mark the adopted color-selection criteria: F444W--F770W $\geq 2$, F444W--F1130W $\geq 5$, F770W--F1130W $\geq 0.5$, and F770W--F1500W $\geq 0.5$. Sources located to the right of these lines are selected. Gray points denote all initial sources. Only sources with more than four photometric bands are considered as candidates. Blue circles represent preliminary YSO candidates (Fflag = 1), that is, sources that meet the selection criteria of only one fitting method in the subsequent SED fitting. Red diamonds indicate candidates with Fflag = 2, that is, sources satisfying the criteria of both fitting methods. Sources excluded by visual inspection are flagged as Sflag = 1 (Green dots).}\label{exfig:cmd} 
\end{extfigure}

\begin{exttable}[h]
    \centering
    \begin{tabular}{cc|cc}
   \toprule
         Criteria & Count & Cflag & Count\\
         \midrule
         F444W $-$ F770W $\geq$ 2 & 385 & 1 & 300 \\
         F444W $-$ F1130W $\geq$ 5 & 185 & 2 & 58 \\
         F770W $-$ F1130W $\geq$ 0.5 & 187 & 3 & 65 \\
         F770W $-$ F1500W $\geq$ 0.5 & 102 & 4 & 62 \\
    \bottomrule
    \end{tabular}
    \caption{Color criteria used to identify YSO candidates. OOnly sources with more than four photometric bands are included in the table. The first column lists the four color-selection criteria, while the second column gives the number of sources that satisfy each criterion. The third column defines the Cflag, which corresponds to the number of color criteria satisfied by a given source, and the fourth column shows the number of sources with the corresponding Cflag value. A total of 485 sources satisfy at least one color-selection criterion (i.e., $\mathrm{Cflag} \geq 1$).}
    \label{extab:1}
\end{exttable}

\begin{extfigure}[t]
\centering
\includegraphics[width=1.0\textwidth]{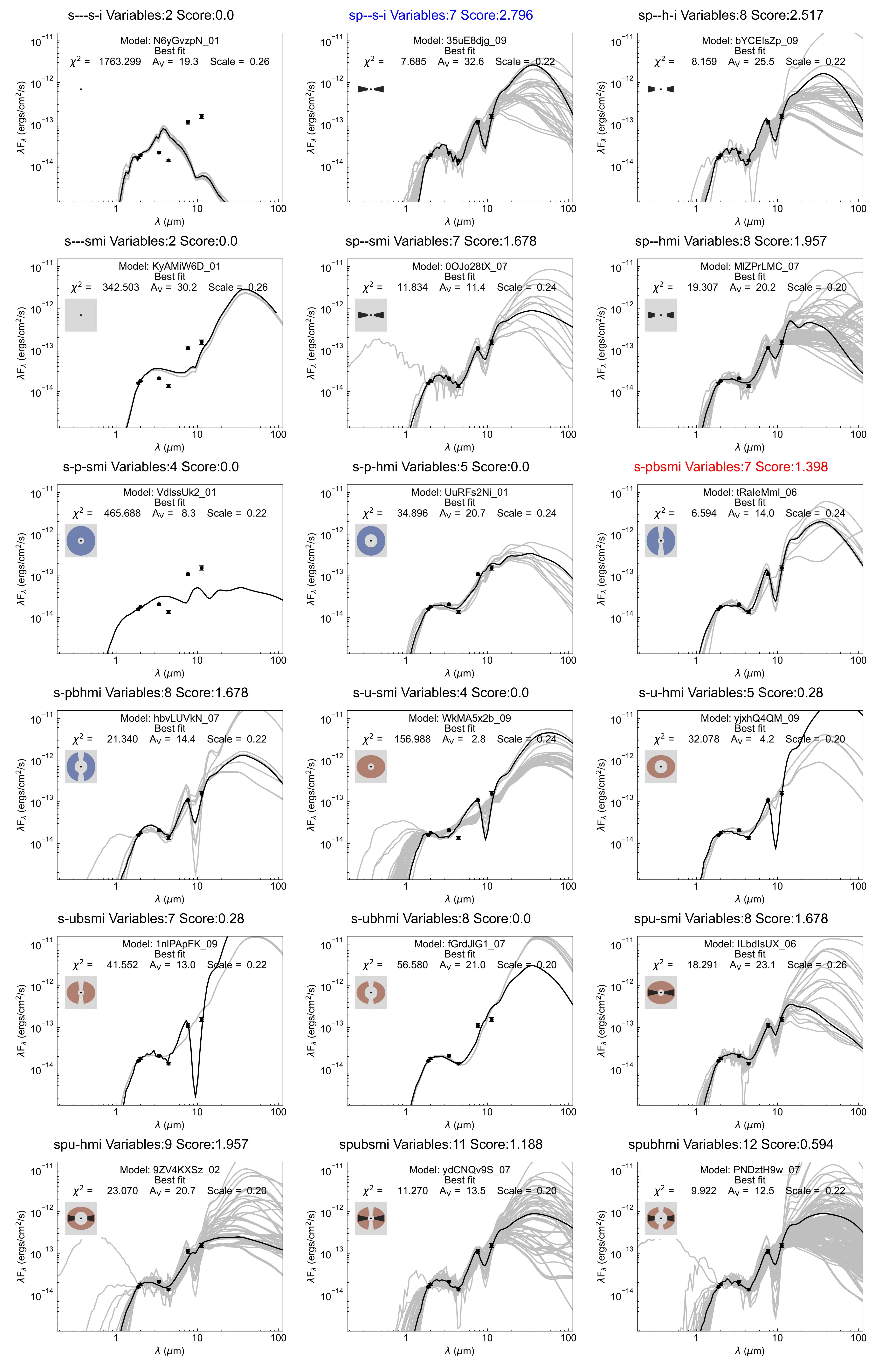}
\caption{An example of the SED fitting for the YSO candidate ``CN26963" using 18 different model sets. The blue titles indicate the model sets selected by the BMC method, while the red titles indicate those selected by the CCF method. A schematic diagram inserted in the upper left corner of each panel (adapted from \cite{2017A&A...600A..11R} under a Creative Commons licence CC BY 4.0) illustrates the composition of the corresponding model set. The title of each panel lists the model set name, the fitting scores, and the number of variables. Black points with error bars represent the observed data, while the black solid line shows the best-fit model from the given model set, and the gray lines indicate all models within that set satisfying $(\chi^2 - \chi^2_{\rm best})/n_{\rm data} < 9$, where $\chi^2_{\rm best}$ is the smallest $\chi^2$ in this set. Note that the models shown here have not been filtered using the ``Above MS" criterion or Equation~3. Therefore, even if some models appear to match the observational data, their fitting scores may still be zero.} \label{exfig:e1}

\end{extfigure}

\begin{extfigure}[t]
  \centering
  \includegraphics[width=0.8\textwidth]{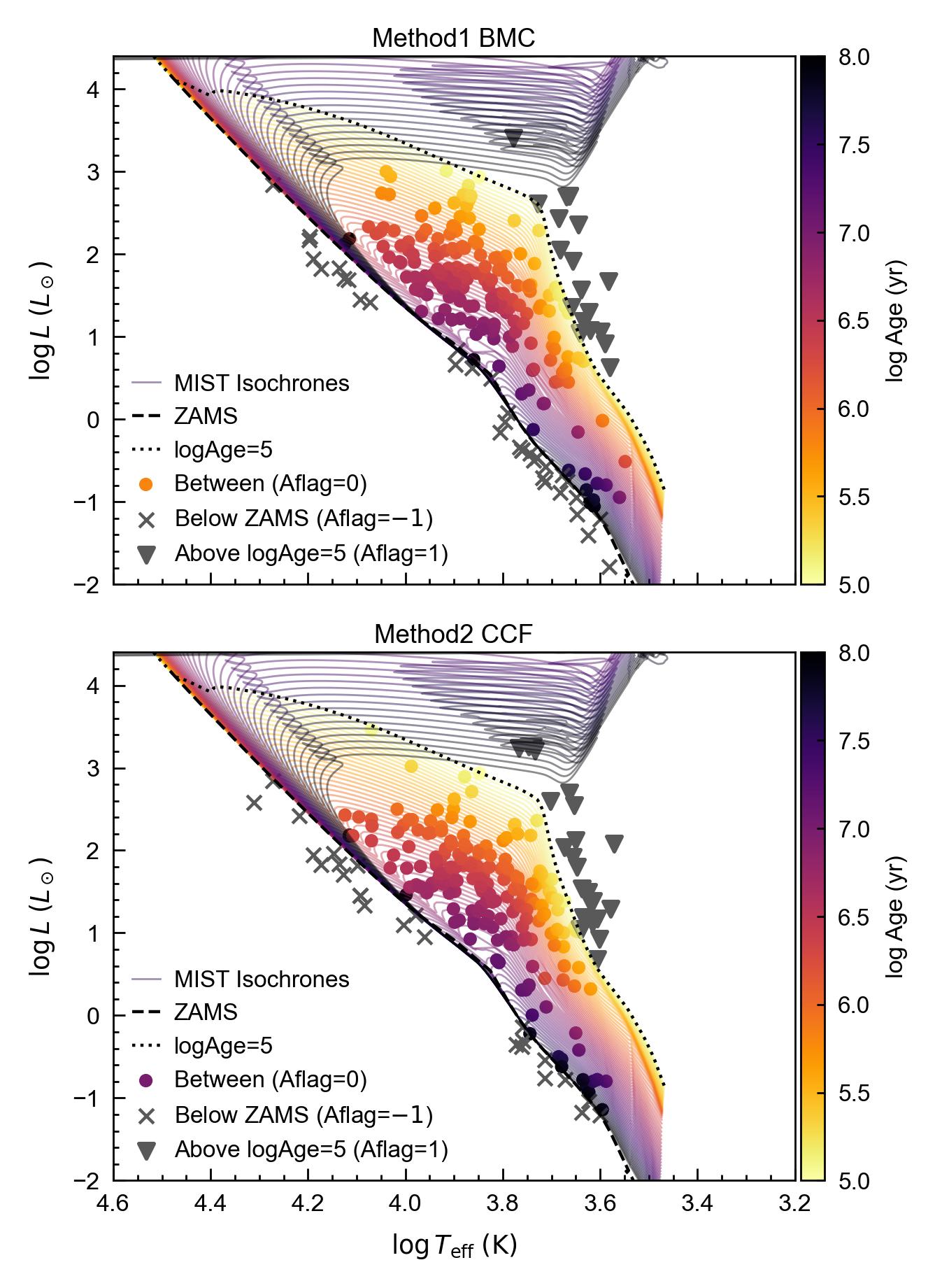} \\

  \caption{Isochrone-fitting results for the two methods. Top: BMC method; bottom: CCF method. Only sources with Fflag = 2 are included, that is, those satisfying the criteria of both fitting methods. The X- and Y-axes represent effective temperature and luminosity, respectively. Each line corresponds to a MIST isochrone: the black dashed line denotes the zero-age main sequence (ZAMS), the black dotted line represents the $\log({\rm Age}) = 5$ isochrone, and the remaining isochrones are color-coded by age, as indicated by the colorbar. Filled circles mark sources located between the ZAMS and the $\log({\rm Age}) = 5$ isochrone, with colors indicating the fitted ages (that is, Aflag = 0). Gray crosses denote sources falling below the ZAMS (that is, Aflag = -1), while gray triangles correspond to sources lying above the $\log{\rm Age}) = 5$ isochrone (that is, Aflag = 1).}
  \label{exfig:am}
\end{extfigure}

\clearpage

\bmhead{Code Availability Statement}
This research made use of Astropy \cite{2013A&A...558A..33A,2018AJ....156..123A,2022ApJ...935..167A} and SEDfitter \cite{2017zndo....235786R} and SAOImage DS9\cite{2000ascl.soft03002S} and the model constructed and developed by \cite{2017A&A...600A..11R,2024ApJ...961..188R}.

\bmhead{Data Availability}
The JWST images used in this work are available from the Mikulski Archive for Space Telescopes (MAST) via program ID 2739, PI: Klaus Pontoppidan. These data can be accessed via the MAST DOI: \url{https://doi.org/10.17909/3w1e-qp71}. The catalog of YSO candidates is provided as source data with this paper.
\bmhead{Competing Interests}
The authors declare no competing interests.
\bmhead{Author Contributions}
BQC and JG developed the initial concept. JW conducted the analysis using JWST catalog which are obtained by JL. JW, BQC and GJ led interpretation of the observational results, aided by MY and BWJ. All authors contributed to manuscript writing and revisions.
\bmhead{Acknowledgements}
This work is supported by the National Natural Science Foundation of China No. 
12322304(BQC), 12173034(BQC), 12133002(BWJ,JG,MY), 12403026(JL) and 12373048(MY), National Natural Science Foundation of Yunnan Province 202301AV070002(BQC) and Xingdian talent support program of Yunnan Province (BQC). We acknowledge the science research grants from the China Manned Space Project with No. CMS-CSST2021-A09(BWJ,JG,BQC), CMS-CSST-2021-A08(BQC) and CMS-CSST-2021-B03(BQC), CMS-CSST-2025-A14(MY). This work is based on observations made with the NASA/ESA/CSA James Webb Space Telescope. The data were obtained from the MAST at the Space Telescope Science Institute, which is operated by the Association of Universities for Research in Astronomy, Inc., under NASA contract NAS 5-03127 for JWST. These observations are associated with program 2739, PI: Klaus Pontoppidan.

\end{document}